\documentclass[letterpaper, 10 pt, conference]{ieeeconf}  

\IEEEoverridecommandlockouts                              

\usepackage{amsmath} 
\usepackage{amssymb}  
\usepackage{color}
\usepackage{graphicx}
\usepackage{epstopdf}
\usepackage{enumerate}   
\usepackage{tikz}

\DeclareGraphicsExtensions{.eps}

\newtheorem{proposition}{Proposition}
\newtheorem{corollary}{Corollary}
\newtheorem{lemma}{Lemma}
\newtheorem{remark}{Remark}

\def\calH{\mathcal{H}}
\def\bp{\mathbf{p}}

\newcommand\copyrighttext{%
  \footnotesize \textcopyright 2017 IEEE. Personal use of this material is permitted.
  Permission from IEEE must be obtained for all other uses, in any current or future 
  media, including reprinting/republishing this material for advertising or promotional 
  purposes, creating new collective works, for resale or redistribution to servers or 
  lists, or reuse of any copyrighted component of this work in other works.}
\newcommand\copyrightnotice{%
\begin{tikzpicture}[remember picture,overlay]
\node[anchor=south,yshift=10pt] at (current page.south) {\fbox{\parbox{\dimexpr\textwidth-\fboxsep-\fboxrule\relax}{\copyrighttext}}};
\end{tikzpicture}%
}

\title{\LARGE \bf
Matched disturbance rejection for energy-shaping controlled underactuated mechanical systems
}

\author{Joel Ferguson$^{1}$, Alejandro Donaire$^{2}$, Romeo Ortega$^{3}$ and Richard H. Middleton$^{1}$
\thanks{$^{1}$Joel Ferguson and Richard H. Middleton are with School of Electrical Engineering and Computing and PRC CDSC, The University of Newcastle, Callaghan, NSW 2308, Australia.
        {\tt\small Email: Joel.Ferguson@uon.edu.au, Richard.Middleton@newcastle.edu.au}}%
\thanks{$^{2}$Alejandro Donaire is with the Department of Electrical Engineering and Information Theory and PRISMA Lab, University of Naples Federico II, Napoli 80125, Italy, and with the School of Electrical Eng. and Comp. Sc. of the Queensland University of Technology, Brisbane, QLD, Australia. 
{\tt\small Email: Alejandro.Donaire@unina.it}}%
\thanks{$^{3}$Romeo Ortega is with Laboratoire des Signaux et Syst\`emes, CNRS-SUPELEC, 91192, Gif-sur-Yvette, France
{\tt\small Email: romeo.ortega@lss.supelec.fr}}%
}

\begin{document}

\maketitle
\copyrightnotice

\begin{abstract}

In this paper, we present a method of applying integral action to enhance the robustness of energy shaping controllers for underactuated mechanical systems with matched disturbances. Previous works on this problem have required a number of technical assumptions to be satisfied, restricting the class of systems for which the proposed solution applies. The design proposed in this paper relaxes some of these technical assumptions.

\end{abstract}

\section{Introduction}

Interconnection and damping assignment passivity-based control (IDA-PBC) is a nonlinear control method whereby the closed-loop system is a passive port-Hamiltonian (pH) system with desired characteristics to comply with the control objectives \cite{Ortega2004}. Many systematic solutions have been proposed for the stabilization of nonlinear systems using IDA-PBC, but the general procedure is still limited by the designers ability to solve the so called \textit{matching equations}. Although the matching equation are difficult to solve in some cases, IDA-PBC has been successful applied to a variety of nonlinear systems such as electrical machines \cite{Petrovic2001,Gonzalez2008}, power converters \cite{Rodriguez2000,Rodriguez2001} and underactuated mechanical systems \cite{Acosta2005}-\cite{Donaire2016a}. In general, the equilibrium of a mechanical system stabilised with IDA-PBC will be shifted when an external disturbance acts on the system. In this paper we are interested in robustifying IDA-PBC {\em vis-\'a-vis} constant external disturbances.

A general design for the addition of integral action to pH systems with the objective of rejecting disturbances was first presented in \cite{Donaire2009} and further discussed in \cite{Ortega2012}. The approach relies on a (possibly implicit) change of coordinates to satisfy the matching equations. The integral action scheme was tailored to fully actuated mechanical systems in \cite{Romero2013a} and underactuated mechanical systems in \cite{Donaire2016}. While in both cases the required change of coordinates to satisfy the matching equations were given explicitly, a number of technical assumption were imposed to do so. In both cases, the proposed integral action controllers were shown to preserve the desired equilibrium of the system, rejecting the effects of an unknown matched disturbance.

More recently, an alternative method for the addition of integral action to pH systems was presented in \cite{Ferguson2015}, \cite{Ferguson}. In these works, the controller is constructed from the open-loop dynamics of the plant. The energy function of the controller is chosen such that it couples the plant and controller states, which allows the matching equations to be satisfied by construction. In addition, the control system studied in \cite{Ferguson} has a physical interpretation and is shown to be equivalent to a control by interconnection (CbI) scheme, another PBC technique \cite{Ortega2007}. The method in \cite{Ferguson2015} was shown to be applicable to mechanical systems with constant mass matrix.

In this paper, we extend the integral action design proposed in \cite{Ferguson2015} to underactuated mechanical systems subject to matched disturbances. The assumption of a constant mass matrix is relaxed, and general mechanical systems are considered. The method proposed in this paper is constructed to directly satisfy the matching equations without the need of the technical assumptions previously used in \cite{Donaire2016}. Specifically, the presented scheme allows the open-loop mass matrix, shaped mass matrix and input mapping matrix to be state dependant. 

\noindent {\bf Notation.} In this paper we use the following notation: Let $x \in\mathbb{R}^n$, $x_1\in\mathbb{R}^m$, $x_2\in\mathbb{R}^s$. For real valued function $\mathcal{H}(x)$, $\nabla\mathcal{H}\triangleq \left(\frac{\partial \mathcal{H}}{\partial x}\right)^\top$. For functions $\mathcal{G}(x_1,x_2)\in\mathbb{R}$, $\nabla_{x_i}\mathcal{G}\triangleq \left(\frac{\partial \mathcal{G}}{\partial x_i}\right)^\top$ where $i \in\{1,2\}$. For fixed elements $x^\star\in \mathbb{R}^n$, we denote $\nabla \mathcal{H}^\star\triangleq \nabla \mathcal{H}(x)|_{x=x^\star}$.
For vector valued functions $\mathcal{C}(x)\in\mathbb{R}^m$, $\nabla_x \mathcal{C}$ denotes the transposed Jacobian matrix $\left(\frac{\partial \mathcal{C}}{\partial x}\right)^\top$. 

\section{Problem Formulation}\label{ProbForm}

In this paper, we consider mechanical systems that have been stabilised using IDA-PBC. This class of systems can be expressed as\footnote{See \cite{Donaire2016} for the detailed explanation and motivation of the problem formulation.}:
\begin{equation}\label{mecdist}
	\begin{split}
		\begin{bmatrix}
	         \dot{q} \\
	         \dot{\bp}
		\end{bmatrix}
		&=
		\underbrace{
		\begin{bmatrix}
	          0_{n\times n} & M^{-1}(q)\mathbf{M}_d(q) \\
	         -\mathbf{M}_d(q)M^{-1}(q) & \mathbf{J}_2(q,\bp)-R_d(q)
		\end{bmatrix}}_{F_m(q,\mathbf{p})}
		\begin{bmatrix}
			\nabla_q \mathbf{H}_d \\ \nabla_\mathbf{p} \mathbf{H}_d
		\end{bmatrix} \\
		&\phantom{---}+
		\underbrace{
		\begin{bmatrix}
		    0_{m\times n} & G^\top(q)
		\end{bmatrix}^\top}_{G_m(q)}
		(u-d) \\
		\mathbf{y}
		&=
		G^\top(q)\nabla_\mathbf{p} \mathbf{H}_d,
	\end{split}
\end{equation}
with Hamiltonian 
\begin{equation}\label{mechOLHam}
\mathbf{H}_d(q,\mathbf{p})= \frac 12 \bp^\top \mathbf{M}_d^{-1}(q) \bp + V_d(q),
\end{equation}
where $q,\mathbf{p} \in \mathbb{R}^n$ are the generalised configuration and momentum vectors respectively, $n$ is the number of degrees of freedom of the system, $u\in\mathbb{R}^m$ is the input, $y\in\mathbb{R}^m$ is the output, $d\in\mathbb{R}^m$ is a constant disturbance, $M(q) > 0$ and $\mathbf{M}_d(q) >0$ are the open-loop and shaped mass matrices of the system respectively, $V_d(q)$ is the shaped potential energy, $G(q)$ is the full-rank input matrix, $R_d (q)= G(q)K_p(q)G^\top(q)$ for some $K_p(q) \geq 0$ is the damping matrix and $\mathbf{J}_2(q,\mathbf{p}) = -\mathbf{J}_2^\top(q,\mathbf{p})$ is a skew-symmetric matrix. We assume that \eqref{mechOLHam} has a strict minimum at the desired operating point $(q,\mathbf{p}) = (q^\star,0_{n\times 1})$. For the remainder of the paper, the explicit state dependency of terms and various mapping are assumed and omitted.

The control objective is to develop a dynamic controller $u=\beta(q,\bp,\zeta)$, where $\zeta \in \mathbb{R}^m$ is the state of the controller, that ensures asymptotic stability of the desired equilibrium  $(q,\bp,\zeta) = (q^{\star},0,\zeta^\star)$, for some $\zeta^\star \in \mathbb{R}^m$, even under the action of constant disturbances $d$.

\section{Previous Work}\label{PrevSol}

A nonlinear PID controller was proposed in \cite{Donaire2016} as a solution to the matched disturbance rejection problem. Under the assumptions:
\begin{itemize}
\item[P.1.] $G$ and $\mathbf{M}_d$ are constant
\item[P.2.] $G^\perp\nabla_q(\bp^\top M^{-1}\bp) = 0_{(n-m)\times 1}$,
\end{itemize}
the control law was proposed to be
\begin{equation}
	\begin{split}
		u = &-\left[ K_pG^\top \mathbf{M}_d^{-1}GK_1G^\top M^{-1} + K_1G^\top \dot{M}^{-1} + K_2K_I \right. \\ 
		&\left.
		\times(K_2^\top + K_3^\top G^\top \mathbf{M}_d^{-1}GK_1)G^\top M^{-1} \right]\nabla V_d \\
		&
		-\left[K_1G^\top M^{-1}\nabla^2V_dM^{-1} + (G^\top G)^{-1}G^\top J_2\mathbf{M}_d^{-1} \right. \\
		&\left.
		+ K_2K_IK_3^\top G^\top \mathbf{M}_d^{-1} \right]\bp \\
		&
		-(K_P G^\top \mathbf{M}_d^{-1} GK_2 + K_3)K_I\zeta \\
		\dot{\zeta}
		=
		&(K_2^\top G^\top M^{-1} + K_3^\top G^\top \mathbf{M}_d^{-1}GK_1G^\top M^{-1})\nabla V_d \\ 
		&+ K_3^\top G^\top \mathbf{M}_d^{-1}\bp,
	\end{split}
\end{equation}
where $K_1 > 0$, $K_P > 0$, $K_I > 0$, $K_3 > 0$ and
\begin{equation}
	K_2 
	= 
	(G^\top \mathbf{M}_d^{-1}G)^{-1}.
\end{equation}
The resulting closed-loop can be expressed as
\begin{equation}\label{DonaireCL}
	\begin{split}
		\begin{bmatrix}
			\dot{q} \\ \dot{z}_2 \\ \dot{\zeta}
		\end{bmatrix}
		&=
		\begin{bmatrix}
			-\Gamma_1 & M^{-1}M_d & -\Gamma_2 \\
			-M_d M^{-1} & -GK_pG^\top & -GK_3 \\
			\Gamma_2^\top & K_3^\top G^\top & -K_3^\top
		\end{bmatrix}
		\nabla H_z \\
		H_z 
		&=
		\frac12 z_2^\top\mathbf{M}_d^{-1}z_2 + V_d(q) + \frac{1}{2}(\zeta-\alpha)^\top K_I(\zeta-\alpha)
	\end{split}
\end{equation}
where,
\begin{equation}\label{Donairez2}
	\begin{split}
		z_2
		&=
		\bp+GK_1G^\top M^{-1}\nabla V_d+GK_2K_I(\zeta-\alpha) \\
		\Gamma_1
		&=
		M^{-1}GK_1G^\top M^{-1} \\
		\Gamma_2
		&=
		M^{-1}GK_2 \\
		\alpha
		&=
		K_I^{-1}(K_p + K_3)^{-1}d.
	\end{split}
\end{equation}
The closed-loop system \eqref{DonaireCL} was shown to have a stable equilibrium at $(q,\bp,\zeta) = (q^\star,0_{n\times 1},\alpha)$. Furthermore, if the output signal
\begin{equation}
	y_{d3}
	=
	\begin{bmatrix}
		G^\top M^{-1}\nabla V_d \\
		G^\top \mathbf{M}_d^{-1}z_2 \\
		K_I(z_3-\alpha)
	\end{bmatrix}
\end{equation}
is detectable, then the equilibrium is asymptotically stable.

The assumptions P.1 and P.2 are necessary to ensure that the dynamics of $z_2$ in \eqref{DonaireCL} match the dynamics of $\mathbf{p}$ in \eqref{mecdist}, using the transformation \eqref{Donairez2}.

\section{Integral Action for Underactuated Mechanical Systems}

In this section we propose an alternative method to add integral action to mechanical systems. This is achieved by first performing a momentum transformation such that the disturbance is pre-multiplied by the identity, rather than $G$. The integral action control law is then defined in the transformed coordinates. The resulting closed-loop is shown to be unique and preserves the desired operating point $q^\star$ of the original system.


\subsection{Momentum transformation}\label{momentumTransform}

To solve the integral action problem, we transform the dynamics \eqref{mecdist} such that the disturbance is pre-multiplied by the identity, rather than $G$. Such a transformation is always possible utilising the following matrix:
\begin{equation}\label{Ttransform}
	T(q)
	=
	\begin{bmatrix}
		\{G^\top G\}^{-1}G^\top \\ G^\perp
	\end{bmatrix},
\end{equation}
where $G^\perp \in \mathbb{R}^{m\times n}$ is a full-rank, left annihilator of $G$. 

\begin{lemma}\label{momLemma}
	Consider the system \eqref{mecdist} under the change of momentum coordinates $p= T\bp$. The dynamics can be equivalently expressed as
	\begin{equation}\label{mecp}
		\begin{split}
			\begin{bmatrix} \dot{q} \\ \dot{p}_1 \\  \dot{p}_2  \end{bmatrix}
			&= 
			\begin{bmatrix} 0_{n \times n} & S_1 & S_2 \\ -S_1^\top  & S_{31}-K_p & S_{32}  \\  -S_2^\top & -S_{32}^\top  & S_{34}  \end{bmatrix}
			\begin{bmatrix}  \nabla_{q}\calH_d \\ \nabla_{p_1}\calH_d \\ \nabla_{p_2}\calH_d \end{bmatrix} \\
			&\phantom{---}
			+
			\begin{bmatrix} 0_{m\times n} & I_{m\times m} & 0_{m\times s} \end{bmatrix}^\top
			(u-d) \\
			y &= \nabla_{p_1}\calH_d \\
			\calH_d&=\frac12 p^\top M_d^{-1}(q) p + V_d(q),
		\end{split}
	\end{equation}
	where $p = \operatorname{col}(p_1,p_2)$, $p_1 \in \mathbb R^m$, $p_2 \in \mathbb R^s$, $s=n-m$,
	\begin{equation}\label{s3}
		\begin{split}
		M_d &= T \mathbf{M}_d T^{\top} \\
		S_1 &= M^{-1}\mathbf{M}_dG\{G^\top G\}^{-1} \\
		S_2 &= M^{-1}\mathbf{M}_dG^{\perp\top} \\
		S_{31} &= \{G^\top G\}^{-1}G^\top J_p G\{G^\top G\}^{-1} \\
		S_{32} &= \{G^\top G\}^{-1}G^\top J_p G^{\perp\top} \\
		S_{34} &= G^{\perp}J_p G^{\perp\top} \\
		\end{split}
	\end{equation}
	and $J_p$ is defined by
	\begin{equation}\label{Jp}
		\begin{split}
		J_p
		&=
		\mathbf{M}_d M^{-1}\nabla_q^\top(T^{-1}p)
				-\nabla_q(T^{-1}p) M^{-1}\mathbf{M}_d \\
				&\phantom{---}+\mathbf J_2(q,T^{-1}p).
		\end{split}
	\end{equation}
	As $J_p = -J_p^\top$, both $S_{31}$ and $S_{34}$ are skew-symmetric.
\end{lemma}

\begin{IEEEproof} The proof of this lemma follows along the lines of the proof of \cite[Lemma 2]{Fujimoto2001a}, \cite[Proposition 1]{Venkatraman2010a} and \cite[Theorem 1]{Duindam208}, therefore the full proof is omitted. An outline of the proof, however, can be found in the Appendix.
\end{IEEEproof}

Importantly, the output of the system under the change of momentum, $y$, remains unchanged. Indeed, 
\begin{equation}\label{outputEquv}
	\begin{split}
		\mathbf{y}
		&=
		G^\top\nabla_\mathbf{p} \mathbf{H}_d \\
		&=
		G^\top T^\top\nabla_{p}\calH_d \\
		&=
		G^\top
		\begin{bmatrix}
			G\{G^\top G\}^{-1} & (G^\perp)^\top
		\end{bmatrix}
		\nabla_{p}\calH_d \\
		&=
		\begin{bmatrix}
			I_{m} & 0_{m\times s}
		\end{bmatrix}
		\nabla_{p}\calH_d \\
		&=
		y.
	\end{split}
\end{equation}

\subsection{Integral action control law}
The integral action control law is now proposed for the underactuated mechanical system described in $(q,p)$ coordinates by \eqref{mecp}.
\begin{proposition}\label{propiacl} Consider the system \eqref{mecp} in closed-loop with the controller 
\begin{subequations}\label{controlLaw}
	\begin{align}
		u &= (-S_{31}+K_p+J_{c_1}-R_{c_1}-R_{c_2})\nabla_{p_1} \calH_d \nonumber \\ 
		&\phantom{---}+ (J_{c_1}-R_{c_1}) \nabla_{p_1}\calH_c \label{iacontroller} \\
		\dot{\zeta}&=-R_{c_2}\nabla_{p_1}\mathcal{H}_d -S_1^\top\nabla_{q}\mathcal{H}_d + S_{32}\nabla_{p_2}\mathcal{H}_d, \label{iazeta}
	\end{align}
\end{subequations}
where 
\begin{equation}
	\mathcal{H}_c = \frac12(p_1-\zeta)^\top K_I(p_1-\zeta),
\end{equation}
$\zeta \in \mathbb{R}^m$, $K_I > 0$ and $J_{c_1}=-J_{c_1}^\top$, $R_{c_1} > 0$ $R_{c_2} > 0$ are constant matrices free to be chosen. Then, the closed-loop dynamics can be written in the pH form,
\begin{equation}\label{iacl}
	\begin{bmatrix}
		\dot{q} \\
		\dot{p}_1 \\
		\dot{p}_2 \\
		\dot{\zeta}
	\end{bmatrix}
	=  
	F(x)
	\begin{bmatrix}
		\nabla_{q}\mathcal{H}_{cl} \\
		\nabla_{p_1}\mathcal{H}_{cl} \\
		\nabla_{p_2}\mathcal{H}_{cl} \\
		\nabla_{ \zeta}\mathcal{H}_{cl}
	\end{bmatrix}
	-
	\begin{bmatrix}
		0_{n\times 1} \\ d \\ 0_{s\times 1} \\ 0_{m\times 1}
	\end{bmatrix},
\end{equation}
where 
\begin{equation}\label{Fx}
	F(x)=  
	\begin{bmatrix} 
		0_{n \times n} & S_1 & S_2 & S_1 \\ 
		-S_1^\top  & J_{c_1}-R_{c_1}-R_{c_2} & S_{32} & -R_{c_2}  \\  
		-S_2^\top & -S_{32}^\top  & S_{34} & -S_{32}^\top \\
		-S_1^\top  & -R_{c_2} & S_{32} & -R_{c_2}  \\  
	\end{bmatrix}
\end{equation}
and $\mathcal{H}_{cl}:\mathbb{R}^{2n+m}\to\mathbb{R}$ is the closed-loop Hamiltonian defined as
\begin{equation}
	\mathcal{H}_{cl}(q,p_1,p_2,\zeta)= \mathcal{H}_d(q,p_1,p_2) + \mathcal{H}_c(p_1,\zeta).
\end{equation}
\end{proposition}

\begin{IEEEproof} 
First notice that $\nabla_{p_1}\mathcal{H}_c = -\nabla_{\zeta}\mathcal{H}_c$. Due to this relationship, the dynamics of $q$ and $p_2$ in \eqref{mecp} are equivalent to the dynamics of $q$ and $p_2$ in \eqref{iacl}.

Considering the dynamics of $\zeta$ in \eqref{mecp} and using $\nabla_{p_1}\mathcal{H}_c = -\nabla_{\zeta}\mathcal{H}_c$ yields
\begin{equation}
	\begin{split}
		\dot{\zeta}
		&=
		-R_{c_2}\nabla_{p_1}\mathcal{H}_d -S_1^\top\nabla_{q}\mathcal{H}_d + S_{32}\nabla_{p_2}\mathcal{H}_d \\
		&=
		-R_{c_2}(\nabla_{p_1}\mathcal{H}_d + \nabla_{p_1}\mathcal{H}_c - \nabla_{p_1}\mathcal{H}_c) -S_1^\top\nabla_{q}\mathcal{H}_d \\
		&\phantom{---}
		+ S_{32}\nabla_{p_2}\mathcal{H}_d \\
		&=
		-R_{c_2}\nabla_{p_1}\mathcal{H}_{cl} -R_{c_2}\nabla_{\zeta}\mathcal{H}_{cl} -S_1^\top\nabla_{q}\mathcal{H}_{cl} + S_{32}\nabla_{p_2}\mathcal{H}_{cl} \\
	\end{split}
\end{equation}
which matches the dynamics of $\zeta$ in \eqref{iacl}.

Finally, considering the dynamics of $p_1$ in \eqref{mecp},
\begin{equation}
	\begin{split}
	 \dot{p}_1
	 &=
	 -S_1^\top\nabla_{q}\calH_d + (S_{31}-K_p)\nabla_{p_1}\calH_d + S_{32}\nabla_{p_2}\calH_d \\
	 &\phantom{---}+ u - d \\
	 &=
	 -S_1^\top\nabla_{q}\calH_d + (J_{c_1}-R_{c_1}-R_{c_2})\nabla_{p_1}\calH_d + S_{32}\nabla_{p_2}\calH_d \\ 
	 &\phantom{---}
	 + (J_{c_1}-R_{c_1}) \nabla_{p_1}\calH_c - d \\
	 &=
	 -S_1^\top\nabla_{q}\calH_d + (J_{c_1}-R_{c_1}-R_{c_2})\nabla_{p_1}\calH_d + S_{32}\nabla_{p_2}\calH_d \\ 
	 &\phantom{---}
	 + (J_{c_1}-R_{c_1}-R_{c_2}) \nabla_{p_1}\calH_c + R_{c_2} \nabla_{p_1}\calH_c - d \\
	 &=
	 -S_1^\top\nabla_{q}\calH_{cl} + (J_{c_1}-R_{c_1}-R_{c_2})\nabla_{p_1}\calH_{cl} \\ 
	 &\phantom{---}
	  + S_{32}\nabla_{p_2}\calH_{cl}- R_{c_2} \nabla_{\zeta}\calH_{cl} - d,
	\end{split}
\end{equation}
which is equivalent to the dynamics of $p_1$ in \eqref{iacl}.

\end{IEEEproof}

\begin{remark}
	In the case that $S_{31}$ and $K_v$ are constant, The choice $J_{c_1} = S_{31}$, $R_{c_1} = K_v$ can be made to simplify the control law \eqref{controlLaw}.
\end{remark}

\subsection{Stability}
For the remainder of this section, the stability properties of the closed-loop system \eqref{iacl} are considered. It is shown that the integral action control \eqref{controlLaw} preserves the desired operating point $q^\star$ of the open-loop system. Further, if the original system is detectable, then the closed-loop system is asymptotically stable.
\begin{lemma}
The closed-loop system \eqref{iacl} has an isolated equilibrium point 
\begin{equation}\label{equilibrium}
	(q,p,\zeta) = (q^\star,0_{n\times 1},-K_I^{-1}(J_{c_1}-R_{c_1})^{-1}d).
\end{equation}
\end{lemma}
\vspace{2mm}
\begin{IEEEproof}
	The dynamics of $q$ in \eqref{iacl} can be simplified to
	\begin{equation}
		\begin{split}
			\dot{q} &= M^{-1}\mathbf{M}_d T^\top\nabla_p\mathcal{H}_d \\
			&=M^{-1}\mathbf{M}_d T^\top M_d^{-1}p.
		\end{split}
	\end{equation}
	As $M, \mathbf{M}_d, M_d, T$ are full-rank, $p=0_{n\times 1}$ and $\nabla_p \mathcal{H}_{d} = 0_{n\times 1}$ at any equilibrium. As $\nabla_{p_2} \mathcal{H}_{cl} = \nabla_{p_2} \mathcal{H}_{d}$,
	\begin{equation}\label{equli3}
		\nabla_{p_2} \mathcal{H}_{cl} = 0_{s\times 1}.
	\end{equation}
	The difference between the dynamics of $p_1$ and $\zeta$ are given by $\dot{p}_1 - \dot{\zeta} = (J_{c_1} - R_{c_1})\nabla_{p_1}\mathcal{H}_{cl} - d$.
	As $\nabla_p\mathcal{H}_d = 0_{n\times 1}$,
	\begin{equation}\label{equli1}
		\nabla_{p_1}\mathcal{H}_{cl} = -\nabla_{\zeta}\mathcal{H}_{cl} = (J_{c_1} - R_{c_1})^{-1}d.
	\end{equation}
	Recalling that $-\nabla_{\zeta}\mathcal{H}_{cl}
			=
			-\nabla_{\zeta}\mathcal{H}_{c}
			=
			K_I(p_1-\zeta)$
	and $p_1 = 0$, \eqref{equli1} can be rearranged to find $\zeta = -K_I^{-1}(J_{c_1}-R_{c_1})^{-1}d$.
	Substituting the equilibrium gradients \eqref{equli3} and \eqref{equli1} into \eqref{iacl} and considering the dynamics of $p$, it results in
	\begin{equation}
		\dot{p} = -\begin{bmatrix} S_1 & S_2 \end{bmatrix}^\top\nabla_q\mathcal{H}_{cl},
	\end{equation}
	which implies that $\nabla_q\mathcal{H}_{cl} = \nabla_q\mathcal{H}_{d} = 0_{n\times 1}$ at any equilibrium as $\begin{bmatrix} S_1 & S_2 \end{bmatrix}$ is full-rank. The equilibrium gradient $\nabla_q\mathcal{H}_{d} = 0_{n\times 1}$ is satisfied by $q = q^\star$.
\end{IEEEproof}

\begin{proposition}\label{propmatched}
	Consider system \eqref{mecp} subject to unknown matched disturbance in closed-loop with the controller \eqref{controlLaw}. The following properties hold:
	\begin{enumerate}[(i)]
	\item The equilibrium \eqref{equilibrium} of the closed-loop system is stable. \label{stability} \\
	\item If the output
	\begin{equation}\label{detectOutput}
		y_{p_1}=\begin{bmatrix} \nabla_{p_1}\mathcal{H}_d \\ \nabla_{p_1} \calH_{c} - (J_{c_1} - R_{c_1})^{-1}d \end{bmatrix}
	\end{equation}
	is detectable, the equilibrium is asymptotically stable. \label{asympStability} \\
	\item If the shaped potential energy $V_d$ is radially unbounded, then the stability properties are global. \label{globStable}
	\end{enumerate}
\end{proposition}

\begin{IEEEproof}
To verify \eqref{stability}, consider the function
\begin{equation}\label{iamatchlyap}
	\mathcal{W} = \mathcal{H}_d(q,p_1,p_2) + \frac12(z - z^\star)^\top K_I(z - z^\star),
\end{equation}
where $z = p_1 - \zeta$ and $z^\star = p_1^\star - \zeta^\star = K_I^{-1}(J_{c_1}-R_{c_1})^{-1}d$, as a Lyapunov candidate for the system. $\mathcal{W}$ has a strict minimum at \eqref{equilibrium} as $\mathcal{H}_d$ is strictly minimised by $(q,p)=(q^\star,0_{n\times 1})$ and $K_I > 0$.

Defining $w = \operatorname{col}(q,p_1,p_2,\zeta)$, the closed-loop dynamics \eqref{iacl} can be equivalently expressed as
\begin{equation}\label{iaclMod}
	\dot w
	=  
	F(x)
	\underbrace{
	\begin{bmatrix}
		\nabla_{q}\mathcal{H}_{cl} \\
		\nabla_{p_1}\mathcal{H}_{cl} - (J_{c_1}-R_{c_1})^{-1}d \\
		\nabla_{p_2}\mathcal{H}_{cl} \\
		\nabla_{ \zeta}\mathcal{H}_{cl} + (J_{c_1}-R_{c_1})^{-1}d
	\end{bmatrix}}_{\nabla_w\mathcal{W}}.
\end{equation}
The equilibrium is stable since $F+F^\top \leq 0$, which implies that $\dot{\mathcal{W}}\leq 0$ along the trajectories of the closed-loop system. 
The claim \eqref{asympStability} follows by considering the structure of $F$ and invoking LaSalle's invariance principle.

Finally, to verify \eqref{globStable}, first note that the component of $\mathcal{W}$ associated with the controller and disturbance, $\frac12(z - z^\star)^\top K_I(z - z^\star)$, is radially unbounded in $z$. Then, recalling that  $\mathcal{H}_d$ is of the form \eqref{mecp} and $M_d^{-1} > 0$, it is clear that $\mathcal{H}_d$ is radially unbounded in $p$. Finally, if $V_d$ is radially unbounded in $q$, then $\mathcal{W}$ is radially unbounded. This implies that the closed-loop system is globally stable.
\end{IEEEproof}

\begin{corollary}\label{CorrDetect}
	If the output of the system \eqref{mecdist} is detectable when $d=0_{m\times 1}$ and $u=0_{m\times 1}$, then the closed-loop system \eqref{iacl} is asymptotically stable.
\end{corollary}

\begin{IEEEproof}
	By Proposition \ref{propmatched}, the equilibrium of the closed-loop \eqref{iacl} is asymptotically stable if $y_{p_1}$ is detectable. The control action \eqref{iacontroller}, evaluated at $y_{p_1} = 0_{2m\times 1}$ is $u = d$. Further, using \eqref{outputEquv}, the output of \eqref{mecdist} resolves to be $\mathbf{y} = y = \nabla_{p_1}\mathcal{H}_d = 0_{m\times 1}$. Substituting $u = d$ and $\mathbf{y} = 0_{m\times 1}$ into \eqref{mecdist} recovers the zero dynamics of the original, undisturbed system. Thus, if \eqref{mecdist} is detectable when $d=0_{m\times 1}$ and $u=0_{m\times 1}$, then the closed-loop system \eqref{iacl} is asymptotically stable.
\end{IEEEproof}

\section{Cart Pendulum Example}
In this section, we apply the presented integral action scheme to the cart pendulum system. For the existing IDA-PBC laws, the shaped mass matrix $\mathbf{M}_d$ is not constant so the integral action scheme of \cite{Donaire2016} cannot be used. 

Stabilisation control of the cart pendulum using IDA-PBC was solved in \cite{Acosta2005}. After partial feedback linearisation, the cart pendulum can be modelled as a pH system of the form
\begin{equation}\label{PendubotOL}
	\begin{split}
		\begin{bmatrix}
			\dot q \\ \dot{\mathbf{p}}
		\end{bmatrix}
		&=
		\begin{bmatrix}
			0_{2\times 2} & I_{2\times 2} \\
			-I_{2\times 2} & 0_{2\times 2}
		\end{bmatrix}
		\nabla \mathcal{H} \\
		&\phantom{---}
		+ 
		\begin{bmatrix}
			0_{2\times 1} \\ \mathbf{G}
		\end{bmatrix}
		\left(u-d\frac{1}{m_c+m_p\sin^2\theta}\right) \\
		\mathcal{H}
		&=
		\frac12 \mathbf{p}^\top M^{-1}\mathbf{p} + \mathcal{V},
	\end{split}
\end{equation}
where $q = \begin{bmatrix} q_1 & q_2 \end{bmatrix}^\top$ is the configuration vector containing the angle of the pendulum from vertical and the horizontal position of the car respectively, $\mathbf{p} = \begin{bmatrix} \mathbf{p}_1 & \mathbf{p}_2 \end{bmatrix}^\top$ is the generalised momenta,
\begin{equation}
	\begin{split}
		M
		&=
		I_{2\times 2} \\
		\mathbf{G}
		&=
		\begin{bmatrix}
			-b\cos(q_1) \\ 1
		\end{bmatrix} \\
		\mathcal{V}
		&=
		a\cos(q_1),
	\end{split}
\end{equation}
$m_c$ and $m_p$ are the masses of the cart and pendulum respectively, $a = \frac{g}{l}$, $b = \frac{1}{l}$, $g$ is the acceleration due to gravity and $l$ is the length of the pendulum. The disturbance $d$ is an unknown constant force collocated with the input $u$. 

Note that the system \eqref{PendubotOL} is not in the form \eqref{mecdist} as the disturbance is not constant. In the remainder of this section, the undisturbed system will be stabilised using IDA-PBC and the resulting closed-loop will be converted into the form \eqref{mecdist} by defining a new input mapping matrix and input.

\subsection{Energy shaping}
In the case that $d = 0_{m\times 1}$, the cart pendulum can be stabilised around a desired equilibrium $(q_1,q_2,\mathbf{p}) = (0,q_{2}^\star,0_{2\times 1})$ using the IDA-PBC law
\begin{equation}\label{PendubotESCtrl}
	\begin{split}
		u &= \{\mathbf{G}^\top \mathbf{G}\}^{-1}\mathbf{G}^\top\{ \nabla_q\mathcal{H} - \mathbf{M}_dM^{-1}\nabla_q\mathbf{H}_d + \mathbf{J}_2\mathbf{M}_d^{-1}\mathbf{p} \} \\
		&\phantom{---}- \frac{1}{(m_c+m_p\sin^2\theta)^2}K_p\mathbf{G}^\top \mathbf{M}_d^{-1}\mathbf{p} + u',
	\end{split}
\end{equation}
where
\begin{equation}
	\begin{split}
		\mathbf{M}_d &=
		\begin{bmatrix}
			\frac{kb^2}{3}\cos^3q_1 & -\frac{kb}{2}\cos^2q_1 \\
			-\frac{kb}{2}\cos^2q_1 & k\cos q_1 + m_{22}^0	
		\end{bmatrix} \\
		V_d
		&=
		\frac{3a}{kb^2cos^2 q_1} 
		+ \frac{P}{2}\left[q_2 - q_{2}^\star + \frac{3}{b}\log\left(\sec q_1 + \tan q_1\right) \right. \\
		&\left.\phantom{---------}
		+ \frac{6m_{22}^0}{kb}\tan^2 q_1\right] \\
		\mathbf{J}_2
		&=
		\mathbf{p}^\top \mathbf{M}_d^{-1}\alpha
		\begin{bmatrix}
			0 & 1 \\
			-1 & 0
		\end{bmatrix} \\
		\alpha
		&=
		\frac{k\gamma_1}{2}\sin q_1
		\begin{bmatrix}
			-b\cos q_1 \\ 1
		\end{bmatrix} \\
		\gamma_1
		&=
		-\frac{kb^2}{6}\cos^3q_1,
	\end{split}
\end{equation}
$P > 0$, $k>0$, $m_{22}^0>0$ are tuning parameters, $K_p > 0$ is a constant used for damping injection and $u'$ is an additional input for further control design.

The cart pendulum \eqref{PendubotOL}, together with the control law \eqref{PendubotESCtrl}, results in the closed-loop
\begin{equation}\label{ESCarkDist}
	\begin{split}
		\begin{bmatrix}
			\dot q \\ \dot{\mathbf{p}}
		\end{bmatrix}
		&=
		\begin{bmatrix}
			0_{2\times 2} & M^{-1}\mathbf{M}_d \\
			-\mathbf{M}_dM^{-1} & \mathbf{J}_2 - GK_p G^\top
		\end{bmatrix}
		\nabla \mathbf{H}_d \\
		&\phantom{---}+ 
		\begin{bmatrix}
			0_{1\times 2} & G^\top
		\end{bmatrix}^\top
		(\tilde u - d) \\
		\mathbf{H}_d
		&=
		\frac12 \mathbf{p}^\top M_d^{-1}(q)\mathbf{p} + V_d(q),
	\end{split}
\end{equation}
where $\tilde u = (M+m\sin^2\theta)u'$, \eqref{ESCarkDist} and $G = \frac{1}{m_c+m_p\sin^2\theta}\mathbf{G}$
Clearly, the closed-loop system is of the form \eqref{mecdist}.

\subsection{Integral action}
Before the integral action control law can be applied, the momentum must be transformed as per Section \ref{momentumTransform}. Taking $G^\perp = (m_c+m_p\sin^2\theta)\begin{bmatrix} 1 & b\cos q_1 \end{bmatrix}$, the necessary momentum transformation is $p = T\mathbf{p}$ with
\begin{equation}
	T(q)
	=
	(m_c+m_p\sin^2\theta)
	\begin{bmatrix}
		\frac{-b\cos q_1}{b^2\cos^2 q_1 + 1} & \frac{1}{b^2\cos^2 q_1 + 1} \\
		1 & b\cos q_1
	\end{bmatrix},
\end{equation}
and results in the transformed Hamiltonian
\begin{equation}
	\calH_d=\frac12 p^\top \underbrace{T^{-\top}\mathbf M_d^{-1}(q) T^{-1}}_{M_d^{-1}(q)}p + V_d(q).
\end{equation}
In the new momentum coordinates, the $S$ matrices can be resolved as per \eqref{s3}:
\begin{equation}
	\begin{split}
		S_1 &=
		\frac{m_c+m_p\sin^2\theta}{b^2\cos^2 q_1 + 1}
		\begin{bmatrix}
			-\frac{kb^3}{3}\cos^4q_1 -\frac{kb}{2}\cos^2q_1 \\
			\frac{kb^2}{2}\cos^3q_1 + k\cos q_1 + m_{22}^0	
		\end{bmatrix} \\
		S_{31}
		&=
		0 \\
		S_{32}
		&=
		\frac{1}{b^2\cos^2 q_1 + 1}
		\begin{bmatrix}
			-b\cos(q_1) \\ 1
		\end{bmatrix}^\top\\
		&\phantom{--}\left[\mathbf{M}_d M^{-1}\nabla_q^\top(T^{-1}(q)p)-\nabla_q(T^{-1}(q)p) M^{-1}\mathbf{M}_d\right.\\
		&\left.\phantom{---} + J_2(q,p)\right]
		\begin{bmatrix}
			1 \\ b\cos(q_1)
		\end{bmatrix}.
	\end{split}
\end{equation}
where
\begin{equation}
	\begin{split}
		J_2
		&=
		p^\top T^{-\top} \mathbf{M}_d^{-1}\alpha
		\begin{bmatrix}
			0 & 1 \\
			-1 & 0
		\end{bmatrix}. \\
	\end{split}
\end{equation}
As $S_{31} = 0$ and $K_p$ is constant, the integral control law \eqref{controlLaw} is simplified by making the selection $R_{c_1} = K_p, J_{c_1} = 0$ which results in
\begin{subequations}\label{CartPendIntLaw}
	\begin{align}
		\tilde u &= -R_{c_2}\nabla_{p_1} \calH_d -K_p \nabla_{p_1}\calH_c \\
		\dot{\zeta}&=-R_{c_2}\nabla_{p_1}\mathcal{H}_d -S_1^\top\nabla_{q}\mathcal{H}_d + S_{32}\nabla_{p_2}\mathcal{H}_d.
	\end{align}
\end{subequations}
As discussed in \cite{Acosta2005}, $V_d$ is radially unbounded on the domain $Q = \left\lbrace(-\frac{\pi}{2},\frac{\pi}{2})\times\mathbb{R}\right\rbrace$ and the system \eqref{ESCarkDist} with $\tilde u = d = 0$ is detectable. Thus, by Proposition \ref{propmatched} and Corollary \ref{CorrDetect}, the closed-loop system is asymptotically stable with region of attraction given by the set $\{Q\times \mathbb{R}^2\times \mathbb{R}\}$.

\subsection{Numerical simulation}
The cart pendulum was simulated using the following plant parameters: $g = 9.8, M = I_{2\times 2}, l = 1, m_c = 1, m_p = 1$. The desired cart position was selected to be $q_2^\star = 0$ and the energy shaping control law \eqref{PendubotESCtrl} was implemented with the controller parameters $k=1, m_{22}^0 = 1, P = 1, K_p = 10$. To reject the effects the disturbance $d$, the control law \eqref{CartPendIntLaw} was applied with the controller storage function $\mathcal{H}_c(p_1,\zeta) = \frac12 K_I(p_1-\zeta)^2$ and $K_I = 0.05$. 

The system was simulated for 60 seconds with state of the plant initialised at $(q_1(0),q_2(0),p(0)) = (0,1,0_{2\times 1})$ and the controller initialised at $\zeta(0) = 0$. For the time interval $t\in[0,30)$ the disturbance was set to $d=0$. At $t=30s$, a disturbance of $d=2$ was applied for the remainder of the simulation.

\begin{figure}[htbp]
	\centering
	\includegraphics[width=0.49\textwidth]{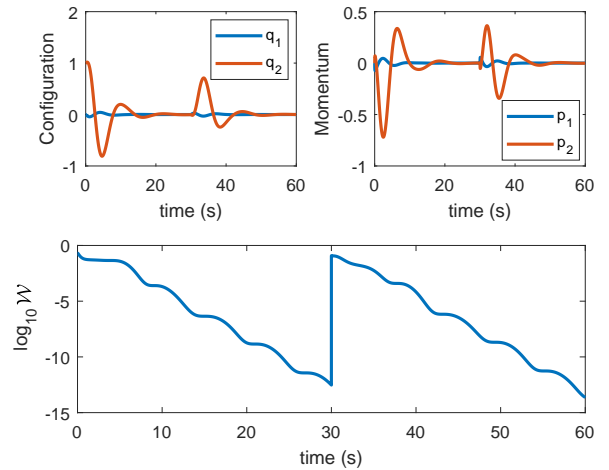}
	\caption{The cart pendulum in closed-loop with an energy shaping controller and integral action subject to a constant disturbance. The system tends toward the desired final position $(q_1,q_2) = (0,0)$ on the interval $t\in[0,30)$. At $t=30$, a disturbance is applied to the system. The integral control compensates for the disturbance and the system tends toward the equilibrium.}
	\label{Cartfig2}
\end{figure}

Figure \ref{Cartfig2} shows that the cart pendulum, together with the integral action control law, tends towards the desired equilibrium on the time interval $t\in[0,30)$. At $t=30$, the disturbance $d=2$ is applied and the states move away from the desired equilibrium. On the time interval $t\in[30,60]$, the integral control compensates for the disturbance and the system again approaches the desired equilibrium.

\section{Conclusions}

In this paper, a method to robustify IDA-PBC via the addition of integral action to underactuated mechanical systems was presented. The method relaxes technical assumptions required by previous solutions. The control scheme preserves the desired equilibrium of the open-loop system, rejecting the effects of an unknown matched disturbance. Further, the closed-loop system was shown to be asymptotically stable provided that the passive output of the open-loop system is detectable.

\addtolength{\textheight}{-4cm}   



\bibliographystyle{IEEEtran}
\bibliography{libraryURLRemoved}           

\appendix\label{app1}

\textit{Proof of Lemma \ref{momLemma}:}
	Let $x_m = \operatorname{col}(q,p)$, $\mathbf{x}_m = \operatorname{col}(q,\mathbf{p})$ and $x_m = g_t(\mathbf{x}_m) = (q,T\mathbf{p})$. The transformed Hamiltonian is defined as
	\begin{equation}
		\begin{split}
			\mathcal{H}_d(q,p)
			&=
			\mathbf{H}_d(q,T^{-1}(q)p) \\
			&=
			\frac 12 p^\top \underbrace{T^{-\top}(q) \mathbf{M}_d^{-1}(q) T^{-1}(q)}_{M_d^{-1}(q)}p + V_d(q).
		\end{split}
	\end{equation}
	Utilising the differential of $g_t$ (see \cite{lee2012introduction}) \eqref{mecdist} can be equivalently expressed in $x_m$ as
	\begin{equation}\label{dynXm}
		\begin{split}
			\dot{x}_m
			&=
			\left\lbrace
			\nabla_{\mathbf{x}_m}^\top g_t F_m \nabla_{\mathbf{x}_m}g_t
			\right\rbrace
			\big|_{\mathbf{x}_m = g_t^{-1}(q,p)} 
			\nabla_{x_m}\mathcal{H}_d \\
			&\phantom{---}+
			\left\lbrace
			\nabla_{\mathbf{x}_m}^\top g_tG_m
			\right\rbrace
			\big|_{\mathbf{x}_m = g_t^{-1}(q,p)} 
			(u-d_m) \\
			&=
			\left\lbrace
			\begin{bmatrix}
				I_{l\times l} & 0_{l\times l} \\
				\nabla_q^\top\left(T\mathbf{p}\right) & T
			\end{bmatrix}
			\begin{bmatrix}
		          0_{l\times l} & M^{-1}\mathbf{M}_d \\
		         -\mathbf{M}_dM^{-1} & \mathbf{J}_2-R_d
			\end{bmatrix} \right. \\
			&\phantom{} \left.
			\times
			\begin{bmatrix}
				I_{l\times l} & \nabla_q\left(T\mathbf{p}\right) \\
				0_{l\times l} & T^\top
			\end{bmatrix}
			\right\rbrace
			\bigg|_{\mathbf{x}_m = g_t^{-1}(q,p)} 
			\begin{bmatrix}
				\nabla_q \mathcal{H}_d \\ \nabla_p \mathcal{H}_d
			\end{bmatrix} \\
			&
			+
			\left\lbrace
			\begin{bmatrix}
				I_{l\times l} & 0_{l\times l} \\
				\nabla_q^\top\left(T\mathbf{p}\right) & T
			\end{bmatrix}
			\begin{bmatrix}
			    0_{l\times m} \\
			    G
			\end{bmatrix}
			\right\rbrace
			\bigg|_{\mathbf{x}_m = g_t^{-1}(q,p)} 
			(u-d_m) \\
			&=
			\begin{bmatrix}
				0_{n\times n} & M^{-1}\mathbf{M}_dT^\top \\
				-T\mathbf{M}_dM^{-1} & T(J_p-R_d)T^\top
			\end{bmatrix}
			\begin{bmatrix}
				\nabla_q \mathcal{H}_d \\ \nabla_p \mathcal{H}_d
			\end{bmatrix} \\
			&\phantom{---}+
			\begin{bmatrix}
				0_{n\times m} \\ TG
			\end{bmatrix}
			(u-d),
		\end{split}
	\end{equation}
	where $J_p$ is defined in \eqref{Jp}. 
	Recalling that $R_d (q)= G(q)K_p(q)G^\top(q)$, the term $TR_dT^\top$ can be simplified to
	\begin{equation}
		\begin{split}
			TR_dT^\top
			&=
		\begin{bmatrix}
			\{G^\top G\}^{-1}G^\top \\ G^\perp
		\end{bmatrix}
		GK_pG^\top
		\begin{bmatrix}
			\{G^\top G\}^{-1}G^\top \\ G^\perp
		\end{bmatrix}^\top \\
		&=
		\begin{bmatrix}
			K_p & 0_{m\times s} \\
			0_{s\times m} & 0_{s\times s}
		\end{bmatrix}.
		\end{split}
	\end{equation}
	Finally, subdividing the momentum variable of \eqref{dynXm} into $p = \operatorname{col}(p_1,p_2)$ and substituting $T$ by its definition \eqref{Ttransform} recovers the dynamics \eqref{mecp}.
	\hspace*{\fill}~\QED\par\endtrivlist\unskip

\end{document}